\documentstyle[12pt,epsf,harvard]{ioplppt}

\begin{document}
\letter{Electron-impact ionization of atomic hydrogen at 2~eV above threshold}
\author{Igor Bray
\thanks{electronic address: I.Bray@flinders.edu.au}
}
\address{
Electronic Structure of Materials Centre,
The Flinders University of South Australia,
G.P.O. Box 2100, Adelaide 5001, Australia}
\date{\today}

\begin{abstract}
The convergent close-coupling method is applied to the calculation of
fully differential cross sections for ionization of atomic hydrogen by
15.6~eV electrons. We find that even at this low energy the method is
able to yield predictive results with small uncertainty. As a
consequence, we suspect that the experimental normalization at this
energy is approximately a factor of two too high.
\end{abstract}
\maketitle

At the base of all electron-atom scattering and ionization problems is 
the fundamental, yet unsolved, three-body problem of an electron
interacting with 
atomic hydrogen. This problem occupies a special place in the set of
unsolved problems of interest to physicists due to its fundamental
nature in the realm of atomic physics. It represents a class of
Coulomb three-body problems which includes electron interaction with
the single positive ion of helium, and hence the problem of helium double
photoionization. 

For heavier atoms the complexity of the Coulomb three-body
problem may be masked by the collective behaviour of the many target
electrons. Similarly, for high incident electron energies the
complicated role played by 
the long-ranged Coulomb interaction is also somewhat hidden. The
problem exhibits all of its complexities at energies a little above
the ionization threshold for the simplest atomic target, namely
hydrogen. Here we have the possibility of exciting a 
countably infinite number of the hydrogen discrete
states as well as the three-body continuum of two very slow
strongly interacting electrons. In this
Letter we consider the e-H problem at the incident electron energy of 15.6~eV, 
i.e. only 2~eV above the ionization threshold.

To solve the e-H problem at a total energy $E$ (presently 2~eV) and
spin $S=0,1$ means
to correctly predict all of the 
possible scattering amplitudes $f^S_{nl}({\bf k})$ for discrete
excitation of 
target eigenstates with energy $\epsilon_{nl}<0$ with
$\epsilon_{nl}+k^2/2=E$, and ionization 
amplitudes $f^S({\bf k}_A,{\bf k}_B)$ with
$k^2_A/2+k^2_B/2=E$. For the discrete transitions the
close-coupling methods have 
proved to be the most successful, particularly at low energies. These
rely on expanding the total wave function in a set of orthonormal
states. From the landmark work of \citeasnoun{YR75}, followed by
\citeasnoun{B78}, 
\citeasnoun{S89} and others, it became clear that the set of orthonormal
states obtained by diagonalising the target Hamiltonian in a Laguerre
basis formed an unusual equivalent-quadrature rule. Thus-obtained
states provide a quadrature rule that incorporates both the 
infinite set of true target discrete states and the true target
continuum. This is an immensely powerful result and forms the basis of
the convergent close-coupling (CCC) method for the calculation of
electron-atom scattering~\cite{BS92,B94,FB95}. The idea relies on
simply increasing the number of expansion states $N=\sum_{l_{\rm
max}}N_l$ until convergence in the amplitude of interest
$f^{SN}_{nl}({\bf k})$ 
is obtained to an acceptable accuracy, just like with standard
numerical quadrature. This approach has
proved very successful for the discrete transitions at all
energies. In the rare case of substantial discrepancy with
experiment~\cite{BS92} subsequent new measurements were found to be in 
agreement with the CCC theory~\cite{YCC97}.

Obtaining reliable scattering amplitudes for the
discrete transitions 
is a good start, but what about ionization? The square-integrable
expansion-states $\phi^N_{nl}$ ($l\le l_{\rm max},n=1,\dots,N_l$),
obtained by diagonalising the target Hamiltonian in a Laguerre basis of 
size $N_l$, have both negative and positive energies
$\epsilon^N_{nl}$. With increasing $N$ the negative-energy states
converge to the true eigenstates ($\epsilon^N_{nl}\to \epsilon_{nl}$,
$\phi^N_{nl}\to \phi_{nl}$), and the
positive-energy states yield an increasingly dense discretization of
the continuum.
By summing the integrated cross sections,
obtained upon solution of the close-coupling equations, for just the
positive-energy states yields excellent
agreement with the measurements of the e-H total ionization cross
section (TICS) \cite{BS93l,KW95,SBBB97}. Though this is the least informative
ionization process it is an encouraging first step. The question is: do the
scattering amplitudes for the excitation of the positive-energy $\phi^N_{nl}$
contain all of the detailed ionization information?

Before proceeding further let us define some convenient
notation. Suppose we are interested in describing an experiment where
the two outgoing electrons have momenta ${\bf k}_A$ and ${\bf k}_B$
with $k_B\le k_A$. When performing the diagonalizations we ensure,
by varying the exponential fall-off parameter $\lambda_l$ \cite{BS92}, that for
each $l$ we have a state $\phi^N_{n_Bl}$ whose energy is
$\epsilon^N_{n_Bl}=k_B^2/2$. We will refer to these states collectively as 
$\phi^N_B$ with energy $\epsilon^N_B=k_B^2/2$. Though it is rarely
practical, let us further suppose that the same diagonalizations have
resulted in states $\phi^N_{n_Al}$ whose energies
$\epsilon^N_{n_Al}=k_A^2/2$. We shall collectively refer to these
states as $\phi^N_A$ with energy $\epsilon^N_A=k_A^2/2$. Similarly, the
scattering amplitudes $f^{SN}_{nl}({\bf k})$ for the excitation of
the states $\phi^N_B$ and $\phi^N_A$, arising upon solution of the
$N$-state close-coupling equations, we write as $f^{SN}_B({\bf k}_A)$ and
$f^{SN}_A({\bf k}_B)$, respectively. Note that the close-coupling formalism
ensures that the total wave function is expanded explicitly
antisymmetrically  using the states
$\phi^N_{nl}$, with the arising equations  
solved separately for each total spin $S$. Thus, each $f^{SN}_{nl}({\bf
k})$ may always be thought of as a combination of the direct $F$ and
exchange $G$ amplitudes,
eg. $f^{SN}_{nl}({\bf k})=F+(-1)^SG$ for hydrogen. The close-coupling
boundary conditions assume 
that only one electron is ever allowed to escape to true infinity,
asymptotically as a plane wave. It is helpful to keep in mind that
the energies $\epsilon^N_B\le\epsilon^N_A$ are symmetrically on either
side of $E/2$, and that the summation of the integrated
cross sections to obtain TICS
includes both sets of amplitudes $f^{SN}_B({\bf k}_A)$ and
$f^{SN}_A({\bf k}_B)$ combined as cross sections. For equal-energy sharing
$\epsilon^N_B=\epsilon^N_A=E/2$, which we consider as the limit
$\epsilon^N_B\to \epsilon^N_A$.

The work of \citeasnoun{BF96} attempted to provide a
correct interpretation of the already calculated positive-energy-state
scattering amplitudes, with some surprising and controversial
results. It was shown that the (e,2e) ionization amplitudes may be defined
from the $f^{SN}_B({\bf k}_A)$ by 
\begin{equation}
f^{SN}({\bf k}_A,{\bf k}_B)=\langle {\bf k}^{(-)}_B|\phi^N_B\rangle
f^{SN}_B({\bf k}_A),
\label{e2eamp}
\end{equation}
where ${\bf k}^{(-)}_B$ is a Coulomb wave (in the case of H target)
of energy $k^2_B/2=\epsilon^N_B$. This definition is in fact a simplification of
the pioneering work of \citeasnoun{CW87}. The overlap has the
effect of changing the unity normalization of $\phi^N_B$ to that of the
true continuum, as well as introducing a one-electron Coulomb phase.
The controversy \cite{BC99}
arises not from the above definition, but from the subsequent use of
(\ref{e2eamp}) to define the triply differential cross section (TDCS) by
\begin{equation}
\frac{d^3\sigma^{SN}({\bf k}_A,{\bf k}_B)}{d\Omega_Ad\Omega_BdE_A}=
|f^{SN}({\bf k}_A,{\bf k}_B)|^2 + |f^{SN}({\bf k}_B,{\bf k}_A)|^2.
\label{TDCS}
\end{equation}
The second term above looks like an exchange term, but it is not.  
The amplitudes $f^{SN}_B({\bf k}_A)$, and hence
$f^{SN}({\bf k}_A,{\bf k}_B)$  are already a coherent combination of
their own direct and exchange amplitudes as determined by $S$. The two terms 
have very different origin.
The amplitudes $f^{SN}_B({\bf k}_A)$ arise from the excitation 
of the states $\phi^N_B$, with the boundary
condition that the ``${\bf k}_A$'' electron exits as a plane wave
totally shielded from the ion by the bound $\phi^N_B$ electron. For
$\epsilon^N_B < k^2_A/2$ this is the physically sound 
shielding approximation, as used in the Born approximation where the
slow electron is modeled by a Coulomb wave and the fast one by the plane
wave. However, the boundary conditions for the amplitude
$f^{SN}_A({\bf k}_B)$ are unphysical (low-energy outgoing plane
wave shielded by a higher energy bound state).
Yet, these two theoretically distinguishable
amplitudes correspond to the same ionization process since
$E=\epsilon^N_A+\epsilon^N_B$. 

From (\ref{TDCS}) we see that close-coupling yields twice as many
amplitudes as we may expect from formal ionization theory. In the
often used language of direct and exchange amplitudes we have two
such pairs $f^{SN}({\bf k}_A,{\bf k}_B)=F_1+(-1)^SG_1$ and 
$f^{SN}({\bf k}_B,{\bf k}_A)=F_2+(-1)^SG_2$, which are very different for
$\epsilon^N_A\ne\epsilon^N_B$. Note, there is no symmetrization relation
between the close-coupling
theory calculated $f^{SN}({\bf k}_A,{\bf k}_B)$ and $f^{SN}({\bf k}_B,{\bf
k}_A)$ as claimed by \citeasnoun{BC99}. In 
forming the TDCS we have $F_iF_i$, $G_iG_i$ and cross terms $G_iF_i$,
generally very different for each $i=1,2$. A careful
numerical study of the problem led to the suggestion that with increasing
$N$ the second term in (\ref{TDCS}) and hence both $F_2$ and $G_2$
converge to zero~\cite{B97l}. This allows for
consistency with formal ionization theory except that the $f^{SN}({\bf
k}_A,{\bf k}_B)$ are obtained only for
$\epsilon^N_B\le\epsilon^N_A$. However, for finite 
$N$ a consistent interpretation (compatible with the definition of
TICS) of the close-coupling
approach to ionization requires the use of both terms. A further
consequence of the numerical study~\cite{B97l} is that the
close-coupling method is unable to obtain convergence to a
satisfactory accuracy in the singly
differential cross section (SDCS) whenever the true SDCS at equal
energy-sharing is
substantial. Nevertheless, it was argued, that if the true SDCS was
known then accurate angle-differential ionization cross sections could still be
predicted. Here we test this claim at just 2~eV above threshold, where 
the SDCS may be reasonably assumed to be approximately flat \cite{REBF97}.

The concept of convergence with increasing 
$N=\sum_{l\le l_{\rm max}}N_l$ involves both the increase of $l_{\rm
max}$ and $N_l$. We performed a series of calculations for various
$N$. The ones presented may be conveniently denoted
by CCC($N_0,l_{\rm max}$) with $N_l=N_0-l$.
To examine the rate of convergence we present two vastly different
calculations CCC(20,5) and CCC(13,4), which require approximately 2Gb
and 500Mb of computer RAM, respectively. 
In both cases the Laguerre
exponential fall-off parameter was $\lambda_l\approx0.6$ with the
variation performed to ensure that for each $l$ there was a state
$\phi^N_{n_Bl}$ with energy
$\epsilon^N_{n_Bl}=\epsilon^N_{n_Al}$=1~eV. In the present 
equal energy-sharing case the two terms in \eref{TDCS} are evaluated using
the same set of amplitudes, assuming a continuous limit of
$\epsilon^N_B\to\epsilon^N_A$. 

The first test of the calculations is the comparison of the total
ionization cross sections (TICS) and its spin asymmetry $A_I$ with the
highly accurate  
measurement \cite{SEG87} of TICS 1.08 ($10^{-17}{\rm cm}^2$) and the
$A_I\approx0.5$ measurements~\cite{Fetal,Cetal}.  The CCC(20,5) and
CCC(13,4) results for the TICS, $A_I$ are 1.18, 0.50 and 0.91, 0.51,
respectively. Thus, we see that both calculations attribute
approximately the correct amount of electron flux to the two spin
ionization channels. The TICS results from other calculations
typically varied around the experimental value. The reliability of various
close-coupling based theories for the calculation of the TICS at low
energies has been discussed in detail by \citeasnoun{SBBB97}. The
difficulty of the problem of obtaining accurate ionization amplitudes
at this energy is indicated by the fact that the total cross section
is more than forty times bigger,  4.7, 4.8 and 4.6$\pm0.1$
($10^{-16}{\rm cm}^2$) respectively for the CCC(20,5), CCC(13,4) and
experiment of \citeasnoun{ZKKS97}.

Next we consider the energy
distribution within the ionization channels, i.e. the SDCS, defined by
\begin{equation}
\frac{d\sigma}{de}^{SN}\!\!\!\!\!\!(e)=\int d\Omega_Ad\Omega_B|f^{SN}({\bf
k}_A,{\bf k}_B)|^2. 
\label{SDCS}
\end{equation}
The TICS $\sigma^{SN}_{\rm I}$ is obtained by performing the
integration
\begin{eqnarray}
\label{ticsE}
\sigma^{SN}_{\rm I}&=&\int_0^E de \frac{d\sigma}{de}^{SN}\!\!\!\!\!\!(e)\\
&=&\int_0^{E/2} de
\left(\frac{d\sigma}{de}^{SN}\!\!\!\!\!\!(e)+
\frac{d\sigma}{de}^{SN}\!\!\!\!\!\!(E-e)\right).
\label{TICS}
\end{eqnarray}
The integral in \eref{ticsE} is equivalent to the sum of the
integrated cross sections for the excitation of the positive-energy states.
The step function hypothesis \cite{B97l} says that the second term
in  \eref{TICS} converges to zero with increasing $N$. The origin of
the two terms in \eref{TDCS} are the two terms in \eref{TICS}. We
think of the second term as numerical ``left-overs'' from an incomplete
convergence with $N$, due to its minor contribution (past 1~eV) to the
TICS. Integration of \eref{TDCS}, for a given secondary energy $e$,
over the angular variables yields the integrand of \eref{TICS}.

\begin{figure}
\hskip2truecm\epsffile{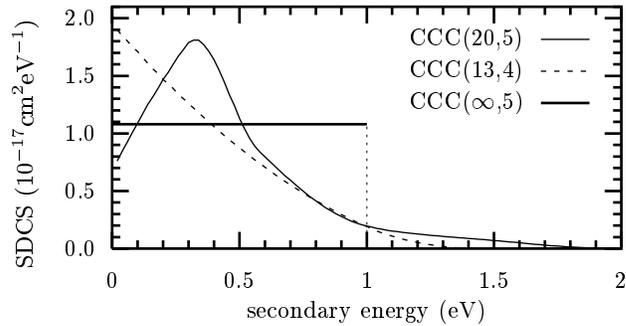}
\caption{The singly differential cross sections arising in the
CCC($N_0,l_{\rm max}$)
(see text) calculations. The step function labeled by CCC($\infty,5$)
is an integral preserving estimate.
}
\label{sdcs}
\end{figure}
In \fref{sdcs} the spin-averaged SDCS are presented. We see that
there is no convergence in the CCC(20,5) and CCC(13,4) results, though
the integral of both is much the same. The step function
CCC($\infty,5$) is an estimate of what the CCC-calculated SDCS would
converge to for $N_l\to\infty$ (there are no problems in obtaining convergence
with increasing $l_{\rm max}$). In other words, we assume that at this
low energy the true SDCS is approximately flat. Since the
close-coupling theory is unitary we cannot have double counting of the 
TICS, and hence suppose that with increasing $N$ the SDCS defined in
\eref{SDCS} 
becomes non-zero only for $0\le e\le E/2$. In experiment the observed SDCS
is symmetric about $E/2$ with the TICS being obtained upon integration 
to $E/2$. Comparison with the experimental SDCS requires both terms of
\eref{TICS}. For the substantially asymmetric energy-sharing kinematics
\begin{figure}
\hspace{-1.0truecm}\epsffile{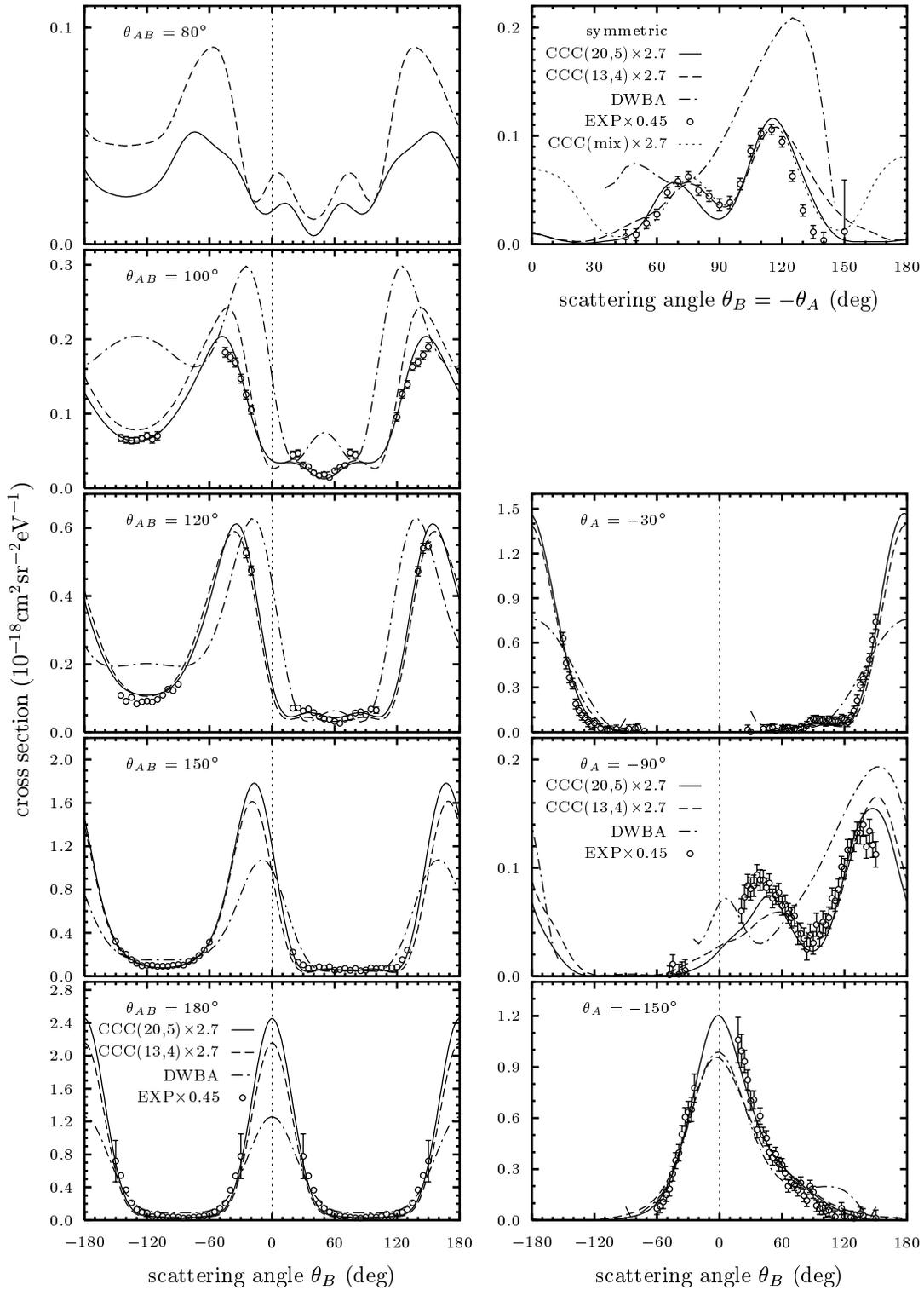}
\caption{The coplanar triply
differential cross sections, in the indicated geometries, for 
electron-impact ionization of atomic hydrogen with 1~eV outgoing
electrons. Absolute experiment of \protect\citeasnoun{Retal97l} has
been scaled by a factor of 0.45 for best 
visual fit to the rescaled CCC data, see text. The DWBA calculations
are due to \protect\citeasnoun{JMS92}.
}
\label{tdcs}
\end{figure}
only the first term contributes significantly, but both are necessary at
equal energy-sharing. From \fref{sdcs} it is clear that the angular
distributions determined by (\ref{TDCS})  will be much too small in
magnitude. In order that the integration of (\ref{TDCS})
over the angular variables, the endpoint of the integrand in
\eref{TICS}, 
yielded the estimated SDCS of 1.08 ($10^{-17}{\rm cm}^2$/eV) we will
multiply the equal energy-sharing CCC-calculated TDCS by
$1.08/(0.2\times2)=2.7$. 

In \fref{tdcs} we present the
TDCS calculated by the two CCC 
models and compare these with experiment and the previously overall
best agreement-yielding theory, the distorted-wave Born approximation
(DWBA) of \citeasnoun{JMS92}. The
relative measurements were initially presented by
\citeasnoun{BBKetal91}, but were remeasured and put on the absolute 
scale, with an estimated 35\% uncertainty, by \citeasnoun{Retal97l}. The
DWBA calculations~\cite{JMS92} work 
relatively well at this low energy since they utilize the effective
charge formalism of \citeasnoun{R68} in the distorting potentials. For 
an example of a more common DWBA approach and the 3C theory see
\citeasnoun{RTL96} and \citeasnoun{BBKetal91}, respectively. 

In the TDCS figure we use the convenient, for the coplanar geometry,
convention that the negative scattering angles are on the opposite side
of the incident beam ($z$-axis). For best visual comparison with the rescaled 
CCC calculations we have multiplied all of the experimental values by the
single constant of 0.45. Having done so, we see excellent agreement
between the two CCC calculations and experiment for all geometries, which is of
considerable improvement on the comparison with the DWBA
calculation. The quality of the agreement gives us confidence that the 
rescaling of the experiment has brought it into consistency with the
estimated SDCS value at 1~eV of 1.08 ($10^{-17}{\rm cm}^2$/eV). Should the
true SDCS prove to be a little convex(concave) then the experimental rescaling
should be done by a factor a little greater(smaller) than 0.45. Perhaps the
experimentally determined normalization is an indication that the SDCS 
is more convex than concave. As a consequence, we do not believe that
the theory of 
Pan and Starace as presented by \citeasnoun{Retal97l} is a factor of
two too low at 15.6~eV, and may indeed be accurate at all
energies. Though not presented they are almost
indistinguishable from the $\theta_{AB}=180^\circ$ rescaled
CCC(20,5) TDCS.

Let us turn specifically to the case where the two detectors are kept
$\theta_{AB}=80^\circ$ apart.
Though no experiment is available for this case we
present it because
it shows the greatest difference between the two CCC calculations, but is
still experimentally measurable. In 
fact, smaller $\theta_{AB}$ geometries yield even greater
differences. Such geometries, first suggested by \citeasnoun{WAW93},
are an excellent test  of the CCC theory
because the cross sections fall rapidly with decreasing
$\theta_{AB}$. We see that the 
bigger calculation yields the smaller cross section for
$\theta_{AB}=80^\circ$. This is an 
important indication of how well the CCC theory is working. For the other
presented cases
the fact that the shapes of the two calculations are much 
the same, even though one requires four times as much
computational 
resources as the other, suggests rapid shape convergence for the
largest cross sections. On the other hand, almost identical
overall magnitude suggests that convergence to the true correct SDCS
is extremely slow.

So how is it that the CCC theory yields such good TDCS angular
distributions? To help answer 
this question let us have a look in more detail at the symmetric geometry.
Given the good agreement between CCC(13,4) and
CCC(20,5) TDCS one would imagine that one may readily interchange the
partial wave amplitudes of (\ref{e2eamp}) $\langle
kl|\phi^N_{nl}\rangle f^{SN}_{nl}({\bf k})$ 
in the two calculations. The curve labeled by CCC(mix) was 
generated by taking the 1~eV $l=1$ partial wave amplitude of the CCC(20,5)
calculation and using it with the other $l$ CCC(13,4)
amplitudes. Whereas one may reasonably expect the CCC(mix) calculated TDCS
to be between the other two, it differs  substantially
when the two electrons emerge close together. This is an indication of 
the importance of treating all partial waves in a consistent
manner. The Laguerre basis choice $N_l=N_0-l$ with similar $\lambda_l$
results in much the same integration rule over the true continuum for
each $l$. In other 
words, the number of positive energy states and their separation is
similar for each $l$. We also use the same set of states for each partial
wave of total orbital angular momentum $J$.
Thus, for each $J$, the error in the energy distribution is
also very similar for each $l$, and this is why the CCC($N_0,l_{\rm
max}$) calculations yield
good TDCS angular distributions whose magnitude is in error by a
single constant. 

What have we learned from this and preceding studies? The CCC
approach to e-H scattering has not fully solved this Coulomb
three-body problem. Given the complexity of the problem it is not
surprising that the close-coupling approach should run into an
intractable problem.
Whereas we are confident of obtaining accurate
discrete scattering amplitudes ab initio at all energies, not so for the
ionization amplitudes. Accurate ionization amplitudes may require too many
states, depending on the incident energy, for practical implementation
of the CCC theory. However, we 
have suggested two empirical prescriptions that still allow for the
CCC-calculated ionization amplitudes to be useful and predictive,
though with some uncertainty. The first,
demonstrated here, ensures
rapid convergence in the angular distributions. It relies on taking a
similar quadrature rule in the continuum for all target-space $l$,
total orbital angular momentum $J$ and total spin $S$. Defining
$N_l=N_0-l$ with $\lambda_l\approx\lambda$ for each $J$ and $S$ achieves
this. There may be other more efficient approaches that use a
different basis as in say the intermediate-energy R-matrix method
\cite{BNS87}. A sensible choice for $\lambda$ is also
important. The second prescription, necessary at low energies when the
true SDCS 
is substantially large at $E/2$, is that of rescaling the cross
sections according to the ratio of the estimated true SDCS and the
close-coupling-calculated SDCS. At high-enough energies no such
rescaling is necessary and most aspects of the problem may be obtained
accurately fully ab initio \cite{BF96,BF96l}.
Given the general structure of the
true SDCS at low energies, and that the close-coupling based theories
obtain the 
correct TICS, estimating the true SDCS is likely to yield only a minor
error. Furthermore, in some cases accurate SDCS are available from
experiment and other theories.
With these two empirical prescriptions the close-coupling approach to
ionization has practical application at all incident energies,
energy-sharing, and geometries of the two detectors.

There are a number of opinions relating to improving the
close-coupling-calculated ionization amplitudes in an ab initio manner
such as by matching the 
calculated total wave function to the correct asymptotic three-body boundary
conditions. However, we note that once the close-coupling equations
have been solved the electron flux has been incorrectly distributed
within the ionization channels. This information will be hidden in the 
total wave function and we suspect will require empirical correction
of some kind prior to matching.
\ack
We thank Steve Jones for providing the DWBA data in electronic form.
Support of the Australian Research Council
and the Flinders University of South Australia is acknowledged.
We are also indebted to the South Australian
Centre for High Performance Computing and Communications.

\end{document}